\documentclass[singlecolumn]{aastex631}

\received{2023 Nov. 30}
\revised{2023 Dec. 26}
\accepted{2023 Dec. 26}
\submitjournal{The Astronomical Journal}

\shorttitle{Gaia21bcv}
\shortauthors{Hodapp et al.}

\begin{document}
\turnoffedit
\title{An Episode of Occultation Events in Gaia21bcv}


\correspondingauthor{Klaus Hodapp}
\email{hodapp@ifa.hawaii.edu}

\author[0000-0003-0786-2140]{Klaus W. Hodapp}
\affil{University of Hawaii, Institute for Astronomy, 640 N. Aohoku Place, Hilo, HI 96720, USA}

\author[0000-0002-5258-6846]{Eric Gaidos}
\affil{Department of Earth Sciences, University of Hawai'i at M\"{a}noa, Honolulu, HI 96822, USA}
	
\author[0000-0002-7064-8270]{Matthew A. Kenworthy}
\affil{Leiden Observatory, Leiden University, P.O. Box 9513, 2300 RA Leiden, The Netherlands}

\author[0000-0002-2471-8442]{Michael Tucker}
\altaffiliation{CCAPP Fellow}
\affiliation{Center for Cosmology and Astroparticle Physics,
The Ohio State University, 
191 West Woodruff Ave,
Columbus, OH, USA}
\affiliation{Department of Astronomy, 
The Ohio State University, 
140 West 18th Avenue,
Columbus, OH, USA}

\author[0000-0003-4631-1149]{Benjamin J. Shappee}
\affil{University of Hawaii, Institute for Astronomy, 2680 Woodlawn Drive, Honolulu, HI 96822, USA}

\author[0000-0003-3490-3243]{Anna V. Payne}
\affil{University of Hawaii, Institute for Astronomy, 2680 Woodlawn Drive, Honolulu, HI 96822, USA}
\affil{Space Telescope Science Institute, 3700 San Martin Drive, Baltimore, MD 21218}

\author[0000-0003-3429-7845]{Aaron Do}
\affil{University of Hawaii, Institute for Astronomy, 2680 Woodlawn Drive, Honolulu, HI 96822, USA}



\begin{abstract}

A previously unremarkable star near the Canis Major OB1/R1 association underwent an episode of multiple deep brightness minima.
Light curves based on archival $Gaia$, ZTF, NEOWISE data and additional observations from LCO and UKIRT show that the star was not variable prior to 2019 Aug 18 (MJD 58700), and on that date started showing brightness dips of up to 3 magnitudes
in the $Gaia$ $G$ and ZTF $r$ bandpasses.
After MJD 59500, $\approx$ 800 days after the onset of these dipping events, the star returned
to its previous brightness, and no significant dipping events have been recorded since.
Compared to the stable phase, NEOWISE infrared photometry in the $W1$ and $W2$ bands indicates a generally redder color, and both decreases and increases in brightness at different times during the dipping episode.
The spectrum of Gaia21bcv taken after the end of the dipping episode shows
several neutral and ionized metal absorption lines, including Li, indicating a spectral
type of $\approx{K5}$.
Variable emission from [\ion{O}{1}] was observed.
The H$\alpha$ absorption in Gaia21bcv is too faint and irregular for this spectral type, indicating that the line is partly filled in by variable emission, a signature of weak episodic accretion. 
Gaia21bcv lies above the zero-age main sequence, but is much fainter than
typical R CrB stars.
We interpret the light curve of Gaia21bcv as being similar to the occultation events in $\epsilon$ Aurigae, i.e., occultation by a disk around
a companion object orbiting the primary star. 
\end{abstract}



\keywords{
Young stellar objects (1834) ---
T Tauri stars (1681) ---
Occulating disks (1149) ---
}


\section{Introduction}

Most young stars are variable, and variability 
is one of the characteristics used by  \citet{Herbig.1962.AdA&A.1.47.TTauri} to define the class of T Tauri stars. 
In addition to the variability caused by modulation of star spots and instabilities in mass accretion onto the star,
dipping events in young stars, i.e., multiple, often quasi-periodic minima,
are due to occultation by dust condensations in the  disks surrounding them, leading to the definition of the UXOr class of 
young variables by
\citet{Herbst.1994.AJ.108.1906.UXOR}.
A summary of the current understanding of dipper objects has recently been published by  \citet{Roggero.2021.AA.651.44.Taurus.Dips}.

In contrast to the dipper phenomenon caused by density inhomogeneities in a circumstellar disk,
where dips are continually occurring,
a small class of variable stars show isolated episodes of deep minima lasting for months to years, 
separated by long phases of constant brightness.
In those cases, the obscuring material must be confined to a small section of its orbit around the star, and these light curves have been interpreted as eclipsing events by a disk surrounding a companion object.

The prototypical object for such substantial periodic occultation events
is $\epsilon$ Aurigae, where the brightness minima are caused by an
optically thick circumstellar disk around a companion star transiting the primary star,
leading to periodic eclipses every 27.1 years
\citep{Hoard.2010.ApJ.714.549.EpsAur}.
Similarly, KH15D discovered by \citet{Kearns.1998.AJ.116.261.KH15D} has in recent decades started showing deep, 
quasi-periodic minima that increased in depth and duration, so that the star no longer reaches the unobscured light level between minima.
The case of the eclipse of 1SWASP J1407 characterized by rapid variations during the occultation event was discussed by
\citet{Kenworthy.2015.ApJ.800.126K.J1407B.rings}
and interpreted as being caused by a giant ring system
around an unobserved object that itself orbits the star.
This scenario of eclipsing disks in general has been discussed by \citet{Mamajek.2012.AJ.143.72.J1407} who concluded that in a sample of 10000 post-accretion stars monitored over 10 years, several such eclipsing events would be predicted.
The Gaia satellite observes any given point in the sky every few months, \citep{Gaia-2016A&A...595A...1G} and the Gaia Alerts project \footnote[1] {\url{http://gsaweb.ast.cam.ac.uk/alerts}} publishes unusual photometric behavior discovered in these repeated observations.
The discovery of Gaia21bcv in the CMa OB1/R1 association of young stars, as a result of the Gaia monitoring, is confirming the prediction by \citet{Mamajek.2012.AJ.143.72.J1407} that more such secondary disk or ring occultation events remain to be discovered.

It should be noted that another class of variable star with similar long-duration drops in brightness are the R Coronae Borealis (RCB) stars, carbon-rich post-main-sequence stars where the condensation of dust clouds is responsible for the dimming of the star.
RCB stars are high-luminosity objects in the late phases of stellar evolution and can be distinguished from other dimming events such as UXors on this basis.

This paper reports the results of a photometric and spectroscopic observing campaign and the analysis of various archival records initiated after the Gaia21bcv alert. 

\section{Observations and Results}

\subsection{Photometry}
\subsubsection{Gaia}

Gaia21bcv was an unremarkable star of constant brightness in all the prior years for which
photometry exists.
Between 2015 and mid-2019, $Gaia$ photometry of this star showed little variation, with a magnitude of G=17.70$\pm$0.03.
On 2019 Aug. 18 (MJD 58713), Gaia recorded the star
at G=20.12, followed by a rapid re-brightening to $\approx$ G=18.6 when Gaia obtained the next light curve data points on 2019 Aug. 31.
The minimum that triggered the Gaia alert on 2021 March 1 (MJD 59457) was actually the third major brightness minimum in the recent episode and was followed immediately by a re-brightening before the object seasonally became unobservable. 
The first few observations of the observing campaign in late 2021, when Gaia21bcv became observable again, recorded the end of another minimum. 
After MJD 59520 (2021 Nov. 2), the object returned to its bright state and no further occultation events were observed.
The Gaia satellite observes any given point in the sky every few months \citep{Gaia-2016A&A...595A...1G}, sampling the rapid dipping events sparsely, but with greater precision than the available ground-based photometry.

\subsubsection{ZTF}
To complement the Gaia light curve data, we have downloaded 
\footnote[2] {https://irsa.ipac.caltech.edu}
archival
$r$-band photometry of Gaia21bcv from the Zwicky Transient Facility (ZTF)
\citep{Bellm.2019.PASP.131.8002.ZTF}
archive \citep{Masci.2019.PASP.131.8003.ZTFarchive}.
We shifted all ZTF magnitudes
by a fixed amount (-0.1 mag) to match the Gaia $G$ photometry.

\subsubsection{NEOWISE}
At the position of Gaia21bcv, the NEOWISE mission \citep{Mainzer.2014.ApJ.792.30.NEOWISE} continues to obtain multiple photometric measurements over about a day, every six months.
For the light curve in Fig.~1, we have median combined all individual measurements from \citet{WISE.2020.archive} in each of these day-long intervals into one measurement. These combined photometric points
and the resulting colors are listed in Table 1.

\begin{deluxetable*}{crrr}
\tabletypesize{\scriptsize}
\tablecaption{Gaia21bcv NEOWISE Photometry\\}
\tablewidth{0pt}
\vspace{0.5cm}
\tablehead{
\colhead{Epoch [MJD]} & \colhead{$W1$ [mag]} & \colhead{$W2$ [mag]} & \colhead{$W1-W2$ [mag]}}
\startdata
55292 & -0.05 $\pm$ 0.03 & -0.02 $\pm$ 0.05 &  0.02 $\pm$ 0.06 \\
55484 & -0.04 $\pm$ 0.03 & -0.03 $\pm$ 0.05 &  0.05 $\pm$ 0.06 \\
56949 &  0.00 $\pm$ 0.03 & -0.02 $\pm$ 0.05 &  0.08 $\pm$ 0.06 \\
57116 &  0.01 $\pm$ 0.03 & -0.05 $\pm$ 0.05 &  0.12 $\pm$ 0.06 \\
57311 &  0.02 $\pm$ 0.03 &  0.05 $\pm$ 0.05 &  0.04 $\pm$ 0.06 \\
57476 &  0.00 $\pm$ 0.03 & -0.01 $\pm$ 0.05 &  0.07 $\pm$ 0.06 \\
57675 & -0.01 $\pm$ 0.03 & -0.04 $\pm$ 0.05 &  0.10 $\pm$ 0.06 \\
57836 &  0.01 $\pm$ 0.03 & -0.03 $\pm$ 0.05 &  0.09 $\pm$ 0.06 \\
58042 & -0.04 $\pm$ 0.03 & -0.07 $\pm$ 0.05 &  0.09 $\pm$ 0.06 \\
58196 & -0.01 $\pm$ 0.03 & -0.02 $\pm$ 0.05 &  0.07 $\pm$ 0.06 \\
58406 & -0.01 $\pm$ 0.03 &  0.06 $\pm$ 0.05 & -0.01 $\pm$ 0.06 \\
58564 &  0.02 $\pm$ 0.03 & -0.03 $\pm$ 0.05 &  0.10 $\pm$ 0.06 \\
58771 & -0.11 $\pm$ 0.03 & -0.22 $\pm$ 0.05 &  0.17 $\pm$ 0.06 \\
58928 &  0.55 $\pm$ 0.03 &  0.27 $\pm$ 0.05 &  0.34 $\pm$ 0.06 \\
59138 &  0.30 $\pm$ 0.03 &  0.09 $\pm$ 0.05 &  0.27 $\pm$ 0.06 \\
59292 & -0.16 $\pm$ 0.03 & -0.41 $\pm$ 0.05 &  0.31 $\pm$ 0.06 \\
59502 &  0.01 $\pm$ 0.03 &  0.03 $\pm$ 0.05 &  0.03 $\pm$ 0.06 \\
59659 &  0.06 $\pm$ 0.03 &  0.07 $\pm$ 0.05 &  0.04 $\pm$ 0.06 \\
59866 &  0.05 $\pm$ 0.03 &  0.10 $\pm$ 0.05 &  0.01 $\pm$ 0.06 \\
\enddata
\end{deluxetable*}

\subsubsection{UKIRT/WFCAM}

We have monitored Gaia21bcv in the $z$, $J$, $H$, and K bands using WFCAM \citep{Casali2007} on UKIRT.
These observations started just when the eclipsing episode was coming to an end and only the first 40 days of this observing campaign contain information about the absorbing material.
After that, the data only confirm that post-eclipse Gaia21bcv has returned to the same brightness as its 2MASS catalog entry.

\subsubsection{Las Cumbres Observatory}

Gaia21bcv was observed over a span of 94 days between MJD 59496.8 and 59591.0 with the 1-m  telescope network of the Las Cumbres Observatory Global Telescope \citep[LCOGT; ][]{Brown.2013.PASP.125.1031.LCO}.  
Photometry was obtained in $Bessell$ $V$, $R$, and $I$ filters with exposure times of 460, 80, and 50 sec, respectively.
Image processing and extraction of instrumental magnitudes were automatically performed using the LCOGT {\tt BANZAI} pipeline \citep{McCully.2018.SPIE.10707E.0KM}, while determination of relative magnitudes corrected for individual image zero-points was carried out by custom software routines \citep{Gaidos.2022.MNRAS.514.1386G}.  
This monitoring started a few weeks earlier than the other monitoring campaigns at the beginning of the visibility period, and the first few data points recorded the last few days of the last dimming event.

\begin{figure}[h]
	\begin{center}
		\includegraphics[angle=0.,scale=0.70]{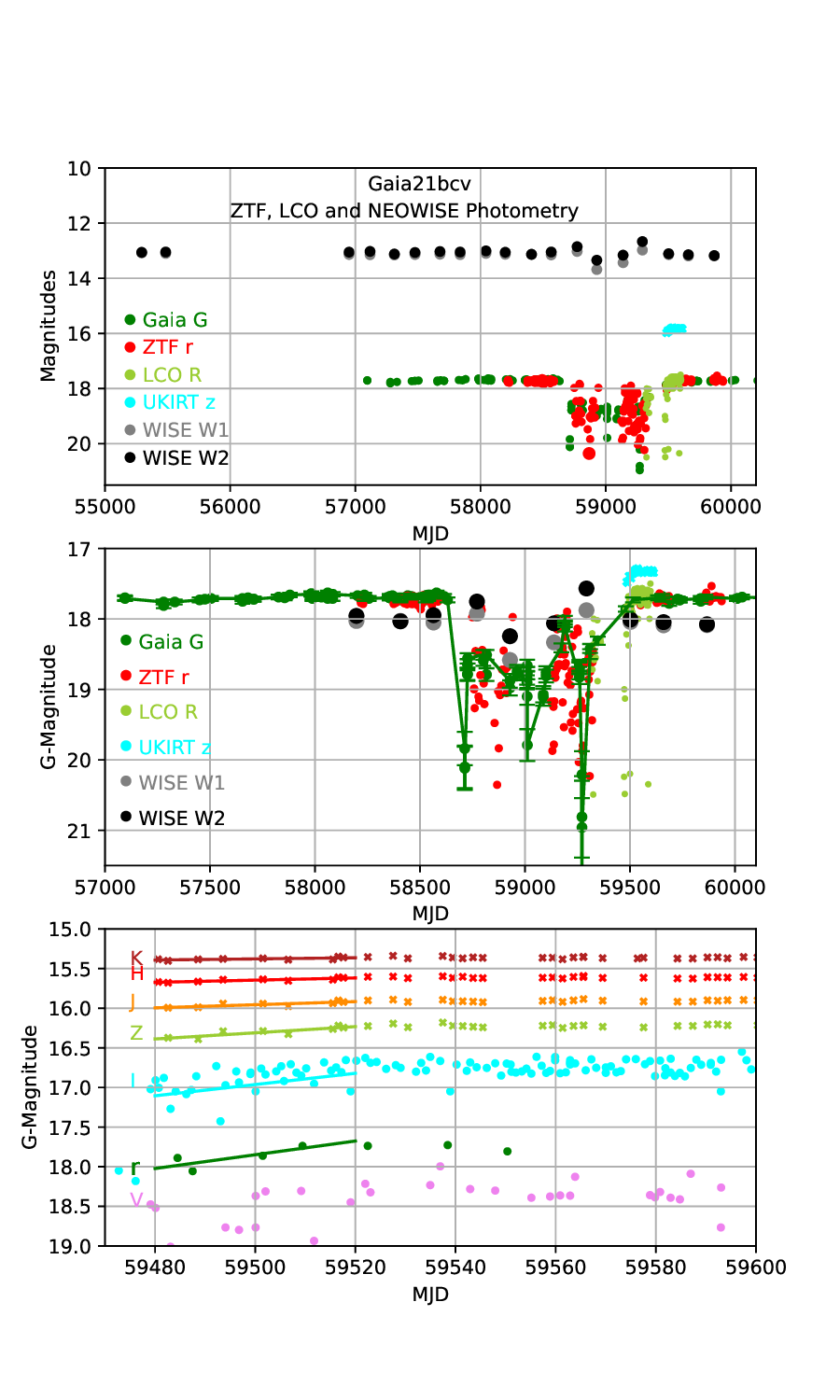}
		\caption{Gaia, ZTF, LCO and NEOWISE light curve. The ZTF (red) data were shifted by 0.1 mag.
                 to match the Gaia $G$ photometry in the stable bright phase. Similarly, the LCO data were calibrated against
                 Gaia $G$ in the bright phase. The errors of the $Gaia$ $G$ photometry are shown, but in the bright phase are smaller than the symbols used. The errors in the ZTF photometry are not shown to preserve some clarity of the figure, but are similar to those of the $Gaia$ data. The errors of the NEOWISE data are not shown in the figure but are listed in Table. 1.
                 The star was essentially constant
			  prior to MJD 58600, after which time several deep dipping events were observed. After
			MJD 59500, these dipping events have ceased and the star has returned to near constant brightness.
                        The top panel gives the overview of the light curve, emphasizing the long time of constant brightness
                        before the dipping episode. The center panel shows details of the dipping episode, with the NEOWISE
                        infrared data shifted by 4.9 mag for clarity.
                        The bottom panel shows the details of the last month of the dipping episode and the return to
                        stable brightness. LCO, ZTF and UKIRT high-cadence photometry is shown, shifted in magnitude for
                        clarity. The first month of the data shown here captured the last of the dipping events and the return to stable brightness. The magnitudes in that last month are fitted by a line to measure the extinction in different filters.
                        The noise in the LCO data is substantially higher than in the infrared UKIRT data, due to the fact that the LCO 1 m telescope is much smaller than the 3.8 m UKIRT, and the star is obscured by interstellar extinction and therefore faint at optical wavelengths.
}
	\end{center}
\end{figure}

\subsubsection{Historic Photometry}

Prior to the Gaia and WISE/NEOWISE missions, there are only a few photometric data points available in various surveys.
The PS1 database \citep{Flewelling.2020.ApJS.251.7.PS1database} photometry of $rPSFMag$ = 17.92 from MJD 56888 (2014 Aug. 19) is consistent with the brightness of Gaia21bcv outside of the dipping episode, considering the differences in the filter bandpasses.
In the infrared, data taken with more consistent filters are available.

The 2MASS magnitudes of Gaia21bcv in the stable bright phase
are $J$ = 14.528, $H$ = 13.676, $K_s$ = 13.338 (on MJD 50904 = 1998 April 1) and UKIDSS ($K = 13.37$ on MJD 55565 = 2011 Jan. 4) \citep{Lawrence.2007.MNRAS.379.1599.UKIDSS}  near-infrared data points are consistent with stable brightness.
The UKIRT WFCAM photometry after the end of the dipping episode 
is $J$ = 14.52, $H$ = 13.723, $K$ = 13.359. This is very close to the 2MASS brightness 
transformed to the UKIRT WFCAM system via the equations in
\citet{Hodgkin.2009.MNRAS.394.675.WFCAM.phot.system} 
of $J$ = 14.4726, $H$ = 13.7356, $K$ = 13.350.
Within the uncertainties of these measurements, 
the brightness after the dipping episode is indistinguishable from that before it.

\subsection{Optical Spectroscopy}
\subsubsection{UH88 Telescope and SNIFS}

We obtained optical spectra of Gaia21bcv at nine epochs between MJD 59502 (2021 Oct. 14) and MJD 59663 (2022 March 25) using the ``Super Nova Integral Field Spectrograph''\citep[SNIFS; ][]{Lantz.2004.SPIE.5249.146L} at the UH 88'' telescope through the SCAT survey \citep{2022PASP..134l4502T}.
The earliest of these observations just recorded the last few days of the last poorly observed minimum in the LCOGT and UKIRT photometry.
The other 8 spectra were obtained when the object had returned to stable brightness.
After MJD 59653, well into the stable bright phase, emission in the [\ion{O}{1}] doublet
at 6300 and 6364 \AA{} was observed, while it was not seen prior to that time.
To improve the S/N over that of the individual exposures, we separately averaged all spectra with
and without the [\ion{O}{1}] line for Fig.~3.

\subsubsection{Keck1/HIRES}
We obtained a high resolution spectrum of Gaia21bcv
on MJD 59617 and 59618 (2022 Feb. 6 and 7),
just after the apparent end of the dipping episode, with the Keck HIRES spectrograph \citep{Vogt.1994.SPIE.2198.362.HIRES}.
We used a fairly narrow slit of 0.6$\farcs$ to achieve a spectral resolution of $\approx$ 60000 and used the longest slit possible to search for emission lines near the star and detected H$\alpha$ and [\ion{S}{2}] emission.
Wavelength calibration used a combination of Th-Ar calibration spectra and telluric OH airglow lines, and \ion{Na}{1} and H$\alpha$ emission in the science data.
The sky and extended background were measured along the slit, away from the point source PSF.
The comparison of telluric \ion{Na}{1} night sky emission and the \ion{Na}{1} absorption in the star gave a spectral shift of 0.828 \AA, corresponding to a 31 $\pm$ 1 km s$^{-1}$ radial velocity relative to the solar system barycenter.
Spectra extracted on-source without sky subtraction are shown as blue lines in Fig.~4, while sky-subtracted spectra are shown in red.

\section {Data Analysis and Results}

\subsection{Location and Distance}

Gaia21bcv at coordinates
108\fdg63865, -12\fdg22426 (J2000.0)
is listed in the Gaia EDR3 catalog
\citep{Gaia-2022-arXiv-DR3} as object number $3045209156636885760$,
with a parallax of 0.72 $\pm$ 0.13 mas and renormalized unit weight error (RUWE) of 1.06,
giving a distance of 1382 pc (1178 - 1672 pc)
and distance modulus 10.70 $mag$.
Gaia21bcv (cyan circle in Fig. 2) appears to lie on the south-eastern edge of a region of reduced star density on DSS red and blue, identified as dark nebula $DOBASHI 5098$ \citep{Dobashi.2011.DarkClouds}.
The PS1/UKIRT $gJK$ color composite image in Fig.~2 shows a large number of very red objects in the area of the dark cloud, which also contains a mid-IR source WISE J071429.04-121239.5 indicated by a yellow circle in Fig. 2.

\citet{Gregorio-Hetem.2021.A&A.654.150.CMaOB1.member} have studied young stars in the Canis Major OB1/R1 OB association that forms a ring-like system of molecular clouds with the brightest optically visible object being the \ion{H}{2} region $Sh 2-296$ near the western edge of the association.
They concluded, based on its parallax and proper motion, that Gaia21bcv is  a Class III object with a probability of membership in the OB association of 85\%.
Numerous H$\alpha$ emitting young stars have been found in the CMa OB1/R1 region by \citet{Pettersson.2019.A&A.630.90.noHa},  including the area immediately surrounding Gaia21bcv, but Gaia21bcv was not included in their list of H$\alpha$ stars.

Using Gaia DR2 data, \citet{Santos-Silva.2021.MNRAS.508.1033.CMa.Gaia} identified groups of stars in CMa OB1/R1 based on their position and proper motion.
Gaia21bcv lies outside these defined groups, and therefore was not assigned an age. 
A small grouping of bright, blueish stars is just west of  Gaia21bcv.
The brightest of these is HD 55902 ($V$ = 8.7), a B9III star with low extinction and a 
parallax of 1.52 mas, $d$ = 656 pc.
These bright stars are in the foreground of Gaia21bcv and not physically close to it.

\begin{figure}[h]
	\begin{center}		

		\includegraphics[angle=0.,scale=0.5]{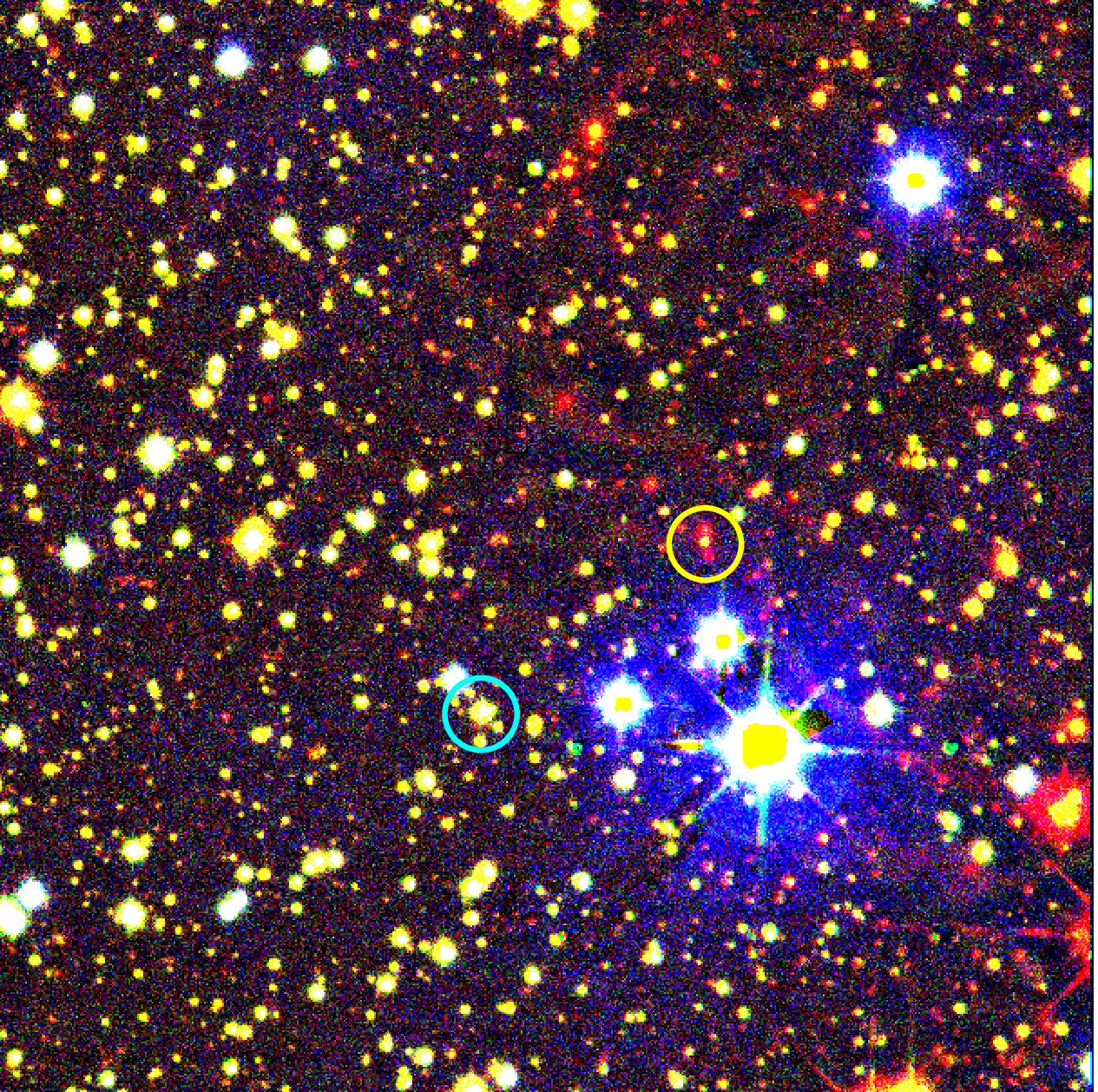}
		\caption{
			gJK color composite images of the $5\arcmin\times5\arcmin$ region near Gaia21bcv, showing extended emission in the
			red image, evidence of absorption. Gaia21bcv is marked by a cyan circle. The yellow circle indicates the position of the embedded infrared source WISE J071429.04-121239.5.
		}
	\end{center}
\end{figure}

The Keck HIRES spectrum detected H$\alpha$ emission both from telluric emission (the narrow emission line) and from the nebulosity near Gaia21bcv (the broad line component). 
Fitting a Gaussian function to the broad line profile gives a barycentric radial velocity of  $27\pm 11$ kms$^{-1}$.

\subsection{Properties of the Star}
\subsubsection{Spectral type}
The combined SNIFS spectra show a continuum slope that in itself would match the continuum slope of a mid-M spectral type. %
However, such a late type can be excluded since we do not observe the deep molecular absorption features, primarily of TiO \citep{Valenti.1998.ApJ.498.851.TiO} characteristic of late spectral types.
The best match to standard spectra is between K4V and K5V just where TiO absorption bands begin to be noticeable.
The comparison K4V and K5V spectra more show pronounced H$\alpha$ absorption, while this line in Gaia21bcv is fainter, and variable in the series of SNIFS medium-resolution spectra.
The steep continuum slope is caused by extinction along the line of sight.
We have dereddened the combined SNIFS spectra to match the overall continuum slope of a  K4-type star by applying the extinction function of \citet{Wang.2019.ApJ.877.116.Extinction} with $A_V$ = 3.2.
The mean wavelength of the $Gaia$ $G$ bandpass is 639.74 nm \citep{Weiler.2018.AA.617A.138W.Gaia.G} for which the \citet{Wang.2019.ApJ.877.116.Extinction} extinction is
A$_G$ = 0.80 A$_V$ = 2.56.
As confirmation of this extinction estimate,
Gaia DR3 has $B_p-R_p$ =2.59 for this star.
The Bayestar19 (Pan-STARRS-based) reddening is $E(B-V)\sim  1.0$, consistent with the extinction value from the spectral slope.


In the top panel in Fig. 3 the bright phase photometry of Gaia21bcv is fitted with a blackbody of $T$ = 4500 K and $A_V$ = 3.2 (red).
To indicate the uncertainty of this combination, $T$ = 4600 K and $A_V$ = 3.0 is indicated in orange and $T$ = 4400 K and $A_V$ = 3.4 in gold.
As confirmation of our spectroscopic determination of the effective temperature, the Gaia DR3 catalog lists the effective temperature as 4520 K, corresponding to a spectral type between K4V and K5V in the tables
\footnote[3] {http://www.pas.rochester.edu/~emamajek/}
by \citet{Pecaut.2013.ApJS.208.9.Teff}.
We adopt a spectral type of K4.5V for Gaia21bcv in the following discussion.
The radius of a K4.5V ZAMS star is 0.707 $R_\odot$, interpolated from the same tables.

In the bright phase, the $K_s$ magnitude of Gaia21bcv in the 2MASS Catalog is 13.338.
The extinction of $A_V$ = 3.2 corresponds to $A_{K_s}$ = 0.238 \citep{Wang.2019.ApJ.877.116.Extinction}. 
With a distance of
1382 pc and a distance modulus of 10.70, Gaia21bcv has an extinction-corrected absolute magnitude of $M_{K_S}$ = 2.40.
With absolute magnitude and colors from the tables of 
\citet{Pecaut.2013.ApJS.208.9.Teff}
a K4.5V star has $M_{K_S}$ = 4.32, making Gaia21bcv 1.92 mag (factor 5.86) brighter than the main sequence and the radius 2.42 times larger, i.e.,
Gaia21bcv has a radius of 1.7 $R_\odot$.
The SED from PS1, 2MASS and WISE data points can be fitted well with a blackbody of T=4500K and extinction
of $A_{V}$ = 3.2 mag and the
\citep{Wang.2019.ApJ.877.116.Extinction}. 
extinction law. Integration of this blackbody fit over the wavelength range gives a luminosity of 4.3 $L_\odot$.
Both arguments clearly indicate that Gaia21bcv is overluminous compared to the ZAMS and therefore a 
pre-main-sequence star.

\begin{figure}[h]
	\begin{center}
		\includegraphics[angle=0.,scale=0.8]{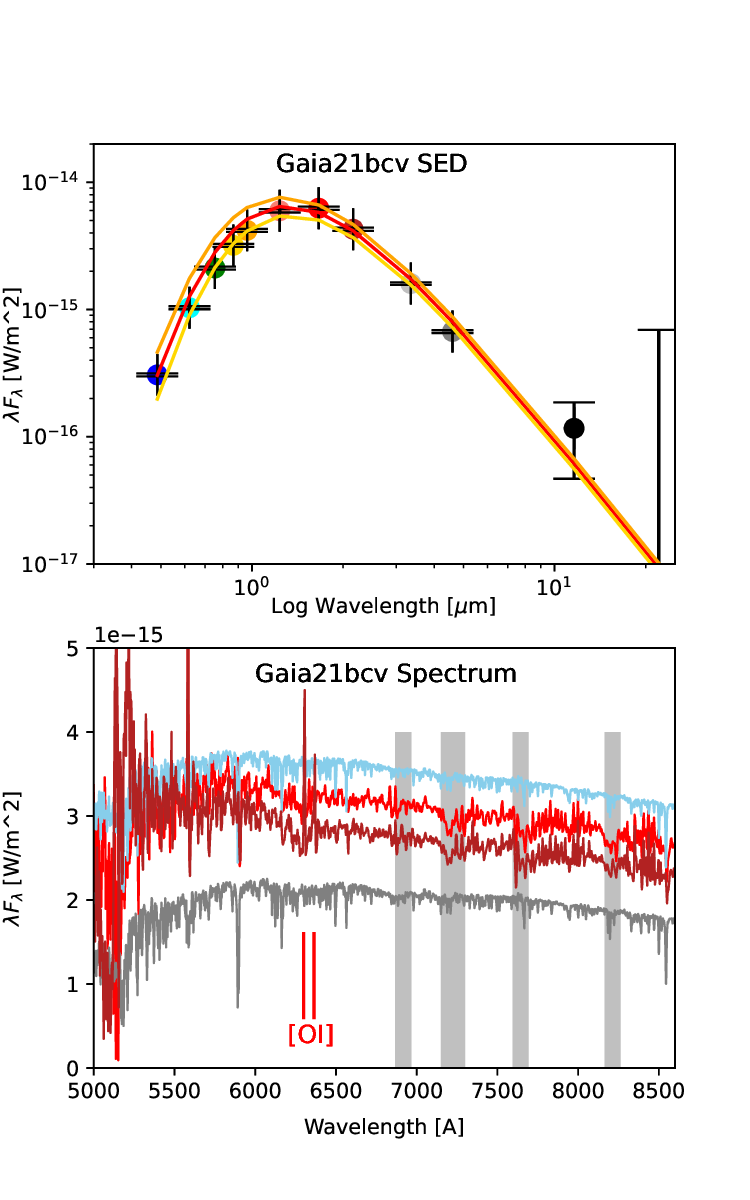}
	
		\caption{
			The top panel shows the photometry of Gaia21bcv in unobscured condition 
			fitted with a blackbody of $T$ =4400 K and $A_V$ = 3.2 (green) or, to indicate the 
			uncertainty of this combination, $T$ = 4600 K and $A_V$ = 3.6. 
			Only the WISE3 data point is substantially above this fit, indicating the presence of
			warm dust.
			The second panel shows the combined SNIFS spectra before (brown) 
			and after (red) the onset of [\ion{O}{1}] emission, dereddened with
			$A_V$ = 3.2. 
			The flux calibrated spectra of K3V (light blue) and K5V (gray) are shown for comparison and were the basis for the spectral 
			classification of Gaia21bcv.
			}
	\end{center}
\end{figure}


\begin{figure}[h]
	\begin{center}
		\includegraphics[angle=0.,scale=0.8]{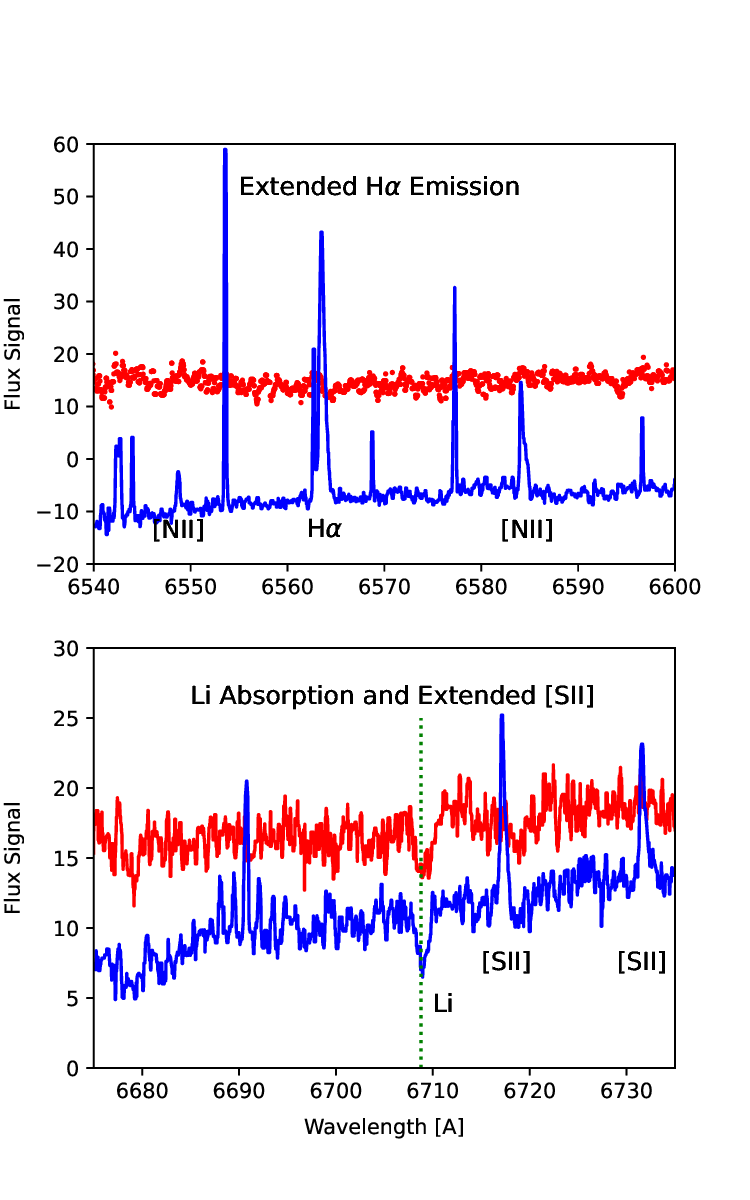}
		\caption{
			The two panels are small sections of the
			Keck HIRES spectrum of Gaia21bcv in stable bright condition, showing absorption and H$\alpha$ and [\ion{N}{2}] extended emission (top) and \ion{Li}{1} absorption in the star (bottom).
   The top panel displays the spectrum without sky subtraction in blue and shows the broad H$\alpha$ and [\ion{N}{2}] lines near Gaia21bcv, and, of course, many narrow telluric night sky emission lines. The sky-subtracted spectrum (red, in different flux scaling for clarity) displays the spectrum of the star itself and does not show any noticeable H$\alpha$ absorption or emission.
   The bottom panel similarly displays the sky-subtracted spectrum of Gaia21bcv in red and the spectrum of the extended emission plus the star in blue. Both spectra show the \ion{Li}{1} absorption
   line in the star. The blue spectrum without sky subtraction shows the [SII] doublet in emission.   
   }
	\end{center}
\end{figure}

The long-slit Keck HIRES spectrum in Fig.~4 shows that H$\alpha$ and [\ion{N}{2}] emission
that permeates the region near Gaia21bcv (blue spectrum), but net emission at the position of the star in the sky-subtracted spectrum is not
detected (red spectrum).
This is consistent with the finding by \citet{Pettersson.2019.A&A.630.90.noHa} in their objective prism search
for H$\alpha$ emission stars in CMa OB1/R1 region that Gaia21bcv was not included in their list of H$\alpha$ emitters. 

The extended H$\alpha$ emission away from the star has two components: A narrow telluric component and the broader component
originating in the nebulosity near Gaia21bcv. This emission is redshifted by 38 km s$^{-1}$ relative to the telluric emission. A Gaussian fit
to its profile gives $\sigma$ = 11 km s$^{-1}$. 
The radial velocity of the star Gaia21bcv of 31 km s$^{-1}$ is therefore well within the velocity dispersion
of the H$\alpha$ emission line. This is consistent with the star being physically located within this HII emission region.
Based on this limited information, we conclude that Gaia21bcv has the characteristics
of a weak-line Class III pre-main-sequence star with a K4-5 photosphere and variable line emission.

\subsubsection{Lithium Absorption}
The Keck HIRES spectrum of Gaia21bcv shows strong absorption in the Li 6708 \AA~line,
stronger than the \ion{Ca}{1} line at 6717 \AA. The equivalent width of the Li 6708 \AA~
absorption line is 334 $\pm$ 40 m\AA.
In stars with strongly convective atmospheres, i.e., spectral classes K and M,
Lithium is destroyed over a timescale of order 10$^8$ years. As specific examples, 
in the K dwarfs in the $\approx$ 600 Myr old Hyades cluster
\citep{Soderblom.1995.AJ.110.729.Li.Hyades},
did not detect any Li absorption. The younger, $\approx$ 70 Myr old
\citep{Soderblom.1995.AJ.110.729.Li.Hyades}, Pleiades cluster
shows Li absorption in K-type stars \citep{Jones.1996.AJ.112.186.Li.Pleiades}, 
with equivalent widths between 30 and 300 m\AA. In the young, about 5 10$^6$ year old, cluster NGC 2264
\citet{Bouvier.2016.AA.590A.78.Li.NGC2264}
found strong Li absorption in K-type stars, with some dependence
of the line equivalent width on the rotation period. In this young cluster, the Li equivalent
widths lie between 500 and 550 m\AA. The line observed in Gaia21bcv lies just above the distribution
of equivalent width for K-type stars in the Pleiades, and is about half of the typical value in
NGC 2264. This suggests an age for Gaia21bcv between those two cases, and probably closer to that
of the Pleiades. As an estimate, we adopt the rounded average age of these two clusters, 40 $\pm$ 10 $Myr$ for Gaia21bcv.

The width of the absorption lines of the \ion{Na}{1} doublets, the strongest absorption lines in the
spectrum, were measured. 
Under the simplistic assumption that vsini $\leq$ FWHM/2., we get vsini $\leq$ 13 km s$^{-1}$. Compared to the rotational velocities measured by
\citep{Jones.1996.AJ.112.186.Li.Pleiades} in the Pleiades, Gaia21bcv is thus a relatively slow rotator,
which correlates with lower abundance of Lithium, as 
first pointed out by \citet{Butler.1987.ApJ.319L.19.Pleiades.rotation}.

\subsubsection{Emission Lines}
The SNIFS spectra can be divided into two groups: those taken up to MJD 59638 (2022 Feb. 28) without
[\ion{O}{1}] emission and those taken in the 10-day period after MJD 59653 (2022 March 15) up to MJD 59663 (2022 March 25) 
that show prominent emission
of the [\ion{O}{1}] doublet. The onset of this emission, happening about 100 days after the end of the
last dipping event is, apparently, not connected
to the end of these dipping events, but may be caused by independent variations in weak accretion
that happens in the innermost regions of a circumstellar disk.

We note that the comparison of the averaged SNIFS spectra with and without
[\ion{O}{1}] also show varying H$\alpha$ absorption. We conclude that the H$\alpha$
absorption of the stellar photosphere is mostly filled in by time-variable chromospheric emission.

On the other hand, the lack of strong and numerous emission lines, e.g. H$\alpha$, Ca-II infrared triplet, Na\,I that are often observed
in classical T Tauri stars rules out strong accretion in this object, and any explanation that the dips
were caused by absorption from accretion funnels 
\citep{Bessolaz.2008.AA.478.155.accretion.funnels}, 
which have been used to explain dipping events in T Tauri stars
\citep{Pouilly.2021.AA.656A.50.accretion.funnel}.

\subsection{Wavelength Dependence of Absorption}
The monitoring campaign initiated in the late 2021 visibility period of Gaia21bcv was
intended to obtain multi-wavelength photometry of the light curve. As it turned out,
only the first $\approx$ 40 days, from MJD 59480 to 59520, recorded the end of the last
dipping event in this recent episode. We work under the assumption that the dipping
episode is caused by absorption by dust, obscuring the star Gaia21bcv. Therefore, the multi-wavelength
light curves gives some information about the wavelength
dependence of the dust absorption. In each of the filters, we computed the slope of a 
linear regression approximation of the light curve (in magnitudes) in the 40 day time interval.
This slope is proportional to the extinction, and Fig.~5 shows the slope value compared to
the extinction law  
\citep{Wang.2019.ApJ.877.116.Extinction}, normalized to match the measured value
in the UKIRT $K$-band. The wavelength dependence of the light curve slope at the end of the dipping episode
closely matches the wavelength dependence of interstellar extinction on which
the 
\citet{Wang.2019.ApJ.877.116.Extinction} law is based. 
This indicates that the dust in the obscuring material contains particles smaller than
the wavelength of the observations and similar in size distribution to the interstellar
medium.
\begin{figure}[h]
\begin{center}
\includegraphics[angle=0.,scale=0.6]{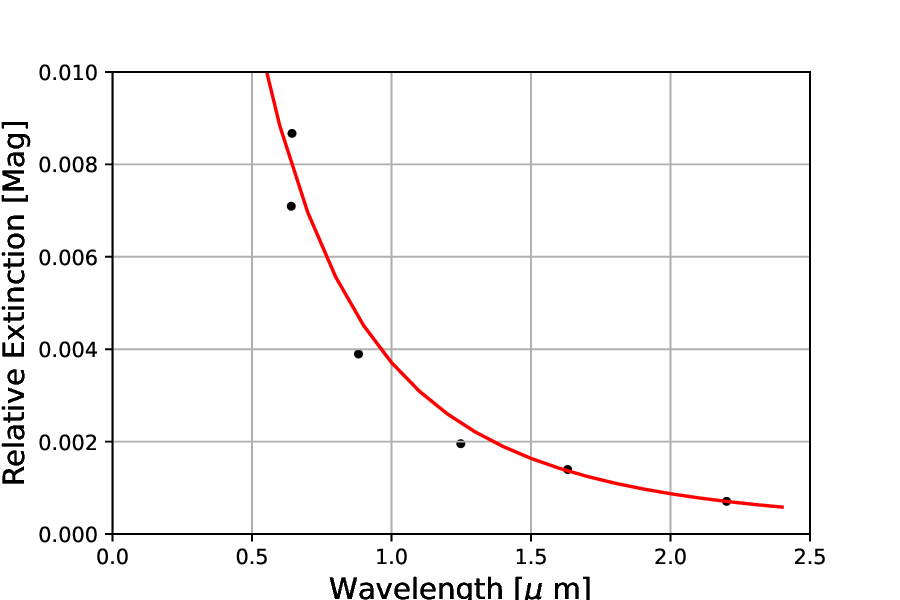}
\caption{The extinction in the last 40 days of the dipping episode have
a wavelength dependence closely resembling interstellar extinction.}
\end{center}
\end{figure}

\subsection{Infrared (WISE) light curve}
The NEOWISE data confirm that Gaia21bcv was of constant brightness prior to MJD 58600.
Four NEOWISE data points were obtained during the dipping episode, during which the $W1-W2$ color was positive (i.e. redder) in contrast to the constant bright phase when the $W1-W2$ color was neutral.
The two data points near the middle of the dipping episode  are fainter than the quiescent brightness, and the other two are brighter.
The brightest $W1$ and $W2$ magnitudes were observed within days of the deepest observed optical dip.

The data points in the stable bright phase, i.e., MJD $\leq$ 58406 and MJD $\geq$ 59659,
have average $W1$ = 13.135 $\pm$ 0.031 (SD) and $W2$ = 13.074 $\pm$ 0.050 (SD), and thus $W1-W2$ = 0.061 $\pm$ 0.016 (standard error of the mean). 
The $A_V$=3.2 mag interstellar extinction of Gai21bcv gives an E($W1-W2$) = -0.042, using the extinction law by \citet{Wang.2019.ApJ.877.116.Extinction}. 
In the list of \citet{Pecaut.2013.ApJS.208.9.Teff} stars in the age range 5-30 Myr
of K4 spectral type have neutral $W1-W2$=0.00 mag color.
The observed $W1-W2$ = 0.06 $\pm$ 0.02 in the stable bright phase
is therefore consistent with a pre-main-sequence K4 star with A$_V$ = 3.2.

During the dipping phase, which we interpret at the result of occultation by some form
of dust cloud we have four NEOWISE data points (Table 1 and Fig.~1). The two data points at the beginning and end of the occultations are below the stable brightness, while the two data points in the middle of the occultation phase are brighter than the unobscured brightness. During this phase, the $W1-W2$ color is redder
at all the four epoch recorded than during the unobscured phase.

\section{Discussion}

For most of the time that we have observations of, Gaia21bcv is of near constant
brightness within the errors of our photometry.
The light curve in Fig.~1 shows a distinct episode of several deep and sharp
minima. The beginning and the end of this episode are equally well defined, and
we do not see evidence for any gradual clearing of the obscuration.
We base this discussion on the assumption that the star is
of constant observed brightness most of the time and that the episode of
dipping events events represents a rare occurrence. 

Based on the wavelength dependence of the absorption (Fig.~1 bottom panel and
Fig.~5) during minima, we conclude the
the dipping events are due to occultations of the star by dust that
contains at least some small particles. We will discuss the nature,
spatial structure, and distance of that dust in the following discussion.

\subsection{Is Gaia21bcv a RCB Star?}
The light curve of Gaia21bcv has similarities to that of some R Coronae Borealis (RCB) variables, 
reviewed by
\citet{Clayton.1996.PASP.108.225.RCBreview}.
In those old, carbon rich AGB supergiants,
episodic formation of carbon dust clouds in the variable wind emerging from the star lead to
episodic, irregular minima lasting months to years that bear a superficial resemblance to those
observed in Gaia21bcv.
For a direct comparison, the prototypical R CrB at a distance of 696 pc and neglecting extinction, has $M_{K_S}$ = -4.65 mag at least, and is therefore clearly in a different luminosity class from Gaia21bcv.
Most RCB stars have spectral types of F and G \citep{Clayton.1996.PASP.108.225.RCBreview}, and their optical spectra show the molecular C$_2$ Swan bands in absorption, characterizing these stars as extremely overabundant in carbon.
The SNIFS spectra of Gaia21bcv show no indication of C$_2$ absorption and are not consistent with a RCB star.
While Gaia21bcv is above the main sequence absolute magnitude, it is far below the luminosity and absolute
V magnitude of R CrB. Its relatively low luminosity, C$_2$-free spectrum, 
and its apparent association with star formation in its vicinity
are the strongest arguments that Gaia21bcv is, in fact, not a RCB variable.

It should be noted that the clear detection of \ion{Li}{1} absorption at nominally 6708 \AA does not strengthen the case
against being a RCB star, since Li can be produced in the He flash forming the RCB star
\citep{Clayton.2011.ApJ.743.44.RCB.Li}.
However, since the RCB explanation is excluded based on the absolute magnitude and lack of C$_2$
absorption, the Li absorption provides additional evidence that Gaia21bcv is indeed 
a young object, independent of the circumstantial evidence from the H$\alpha$ emission, dark clouds
and IRAS sources in its vicinity. 
Main sequence stars with spectral type K and M show very little Lithium absorption as was
already pointed out by \citet{Herbig.1965.ApJ.141.588.Lithium} due to the deep convection in their
atmospheres during their contraction phase and also in their early main-sequence phase that leads to
rapid destruction of Li. The detection of Li in Gaia21bcv with its late spectral type is therefore
another strong indication of a very young star.

\subsection{Occultation by an Orbiting object}

Many young stars show the ''dipping'' behavior, i.e., multiple, repeated
short dips in the brightness. 
The most common dipper objects are very young, still accreting
classical T Tauri stars, of which up to 25\% show dipper-like variability.
For these accreting stars, \citet{Ansdell.2016.ApJ.816.69.dippers} has shown a correlation between
dipping depth and infrared excess, i.e., the more warm dust is present
in the immediate vicinity of the star, the deeper the dips in brightness are.
However, in Gaia21bcv, we do not observe substantial excess emission at 4.6 $\mu$m 
in the bright phase  SED and the bright phase $W1-W2$ color is, 
within the errors, as expected for a K4-5
pre-main-sequence star with A$_V$ = 3.2 mag foreground interstellar extinction.
The deep occultation events and, at the same time, little infrared excess
at near to mid-infrared wavelengths support what is already strongly
suggested by the episodic nature of the occultations and long intervals between these
episodes: The obscuring dust is
concentrated on a small section of the orbit, the overall mass of dust
around the star is small, and the distance of the dust from the star is
so large that, in combination of these factors, infrared excess is not observed at short and mid-infrared wavelengths.
Such a dust cloud at substantial distance from the star could have been produced
by a catastrophic collision of two small planetesimals or asteroids.
Such events have indeed been observed by an increase in infrared emission, which is
approximately isotropic and can be observed from any direction, making the discovery
of such spikes in infrared emission likely. Results on two objects, ID8 and P1121
have been summarized by \citet{Su.2019.AJ.157.202.debris.disk.variability}.
In both objects, orbital periods, collisional cascade build-up time of the dust cloud and
cloud dispersal times are of the order of years.

In at least one case, a recently produced dust cloud has been directly
imaged: Formalhaut b, initially thought to be an exoplanet, was found by \citet{Gaspar.2020.PNAS.117.9712.Formalhaut.b} to be an expanding, very low mass cloud
of debris after what appears to have been an asteroid collision. Based on measurements
of the cloud expansion, they estimate that the cloud had formed only a few years prior
to its discovery and is dissipating on timescales of a few decades. The dissipation time
depends, of course, on the initial velocity of the collision debris.

By comparison, the occulting object around Gaia21bcv is larger and more opaque, but without
a massive object, would probably still dissipate on timescales of decades or centuries.
While we cannot categorically exclude the possibility of a short-lived dust cloud, 
it would be a very unlikely event to
find such a short-lived object at large distances from its host star in just the right orientation
towards the observer to appear transiting in front of the star.

Frequent, sporadic, short, and comparatively shallow dips in the light curve of 
the main sequence star KIC 8462852
were discovered by
\citep{Boyajian.2016.MNRAS.457.3988.dips}. The best explanation for
these short dipping events are swarms of comets briefly
obscuring the star. However, the much longer and deeper brightness minima of Gaia21bcv 
are much more substantial than what could be explained with anything close to
comets in our own solar system.

In addition, the occulting object in Gaia21bcv appears to be well defined, with no
indications of a gradual temporal clearing or smooth profile of the dipping events in the light curve.
The light curve does, instead, suggest a well-defined outer edge of a disk or a ring system.

We therefore favor a model, similar to $\epsilon$ Aur, of a long-lasting disk or ring system
where the containment of the dust or debris cloud along its orbital path must be aided by
the presence of a companion body of dynamically significant mass.

\subsection{Absorption and Scattering}
In the case of Gaia21bcv, the only available data in the mid-infrared are
from the WISE/NEOWISE \citep{Wright.2010.AJ.140.1868.WISE, Mainzer.2014.ApJ.792.30.NEOWISE} mission. The WISE $W3$ band shows a faint, 4.7$\sigma$ detection. The WISE $W4$ image shows a 2-$\sigma$ source that is extended and includes the position of Gaia21bcv.  Based on the same WISE data
\citet{Gregorio-Hetem.2021.A&A.654.150.CMaOB1.member}
have classified Gaia21bcv as a Class III object, i.e., a pre-main-sequence
star without significant infrared excess.

While the $W1-W2$ colors are nearly neutral (zero) in the bright phase, during the
dipping episode, the color was redder ($W2$ brighter). 
There are two possible explanations, not necessarily mutually exclusive,  for
this change in mid-IR colors: 
All dust size distributions containing particles much smaller than the wavelength observed tend to
absorb stronger at shorter wavelengths, explaining a reddening during the occultations.
We have measured the color dependence of the dust absorption in the last few weeks of the
occultation phase, and confirmed a wavelength dependence similar to that of the
interstellar medium, specifically the extinction law by \citet{Wang.2019.ApJ.877.116.Extinction}.

However, somewhat surprisingly, two of the NEOWISE infrared data points during the occultation phase are above the unobscured brightness,
indicating the addition of mid-infrared flux near the middle of this phase.

Any additional thermal emission from the obscuring cloud would be radiated
isotropically and would be observable in all phases of the cloud's orbit. We do not observe any fluctuations of the
$W1$ or $W2$ flux in the unobscured bright phase and therefore, increases in thermal emission must be very rare.
The sudden addition of thermal emission just in the middle of the occultation phase
would be purely coincidental and very unlikely. We therefore suggest that the infrared flux during the occultation is the superposition of two opposite effects: Enhanced flux due to forward scattering in the dust cloud, and, of course, absorption by this same dust. At the optical wavelengths
where most of data were obtained, absorption dominates and no data point during that phase exceeds the unobscured flux. At the WISE $W1$ and $W2$ wavelengths, absorption is present and leads to a reddening of the observed color in all four data points, and the minima at two epochs, but forward scattering directs so much additional flux towards the observer that the center two data points actually exceed
the unobstructed brightness level.

Strong forward scattering has been observed in the more tenuous rings of the Saturn ring system and discussed by \citet{Hedman.2015.Ap.811.67.saturn.rings} as a possible analog
to the dust in debris disks around other stars. Even earlier in the evolution of dust particles, strong scattering at infrared wavelengths is evident in the core shine phenomenon observed in isolated small molecular clouds illuminated by the interstellar
radiation field \citep{Steinacker.2015.AA.582.70.coreshine}.
A detailed analysis or modeling of the dust properties in the Gaia21bcv occulting disk
is beyond the scope of this paper, but based on the properties of both younger and older
dust mixtures mentioned above, strong forward scattering in the occulting disk is a
plausible scenario.

\subsection{A Transiting Ring System}

The model of an orbiting dust cloud, held together by some massive body, is conceptually similar to the now well established model for the periodic deep minima
of $\epsilon$ Aurigae, summarized by 
\citet{Hoard.2010.ApJ.714.549.EpsAur}. 
In this model of $\epsilon$ Aurigae, a dense dust cloud orbiting a companion star occults the primary star periodically, 
but that companion is not itself visible at optical wavelengths.
In contrast to $\epsilon$ Aurigae, the occultation event in Gaia21bcv is not simply a smooth minimum, 
but is a series of sharp minima between brighter periods that in some cases return almost to the
unocculted brightness. 
This indicates more structure in the orbiting dust cloud than is observed in $\epsilon$ Aurigae.
For an eclipse by a disk with sub-structure, for example a ring system, the modulation by
the orbital motion in front of the star is dominant over temporal variations in the
cloud for plausible combinations of orbital period, size of the dust cloud, and rotation
of the dust cloud.

\citet{Mamajek.2012.AJ.143.72.J1407} specifically apply
this model of a highly structured disk around an orbiting companion object,
possibly a ring system, to the case of
SWASP J140747.93-394542.6, where multiple deep minima
spread over $\approx$ 54 days were observed.

Their preliminary model was refined further by 
\citet{vanWerkhoven.2014.MNRAS.4412845.EXORINGS} and
\citet{Kenworthy.2015.ApJ.800.126K.J1407B.rings}
who fit their light curve by an elaborate model of an eclipse by a multiple ring system around 
a planetary companion.

Similarly and even earlier, \citet{Mikolajewski.1999.MNRAS.303.521.EECep}
have discussed the case of EE Cep, where multiple, long-duration minima have been observed, interpreted as
eclipses by an optically thick disk around a companion object.

We have no evidence from the high resolution spectrum of line splitting and therefore the presence
of a stellar companion. Gaia21bcv is listed in the Gaia EDR3 with a RUWE of 1.06, indicating a solid detection with no evidence astrometric wobble that might indicate
an orbiting companion. We can only speculate that the companion object stabilizing the dust cloud  may be an otherwise unobservable companion 
of stellar, sub-stellar, or planetary mass. 

\begin{figure*}[h]
	\begin{center}
		\includegraphics[angle=0.,scale=0.70]{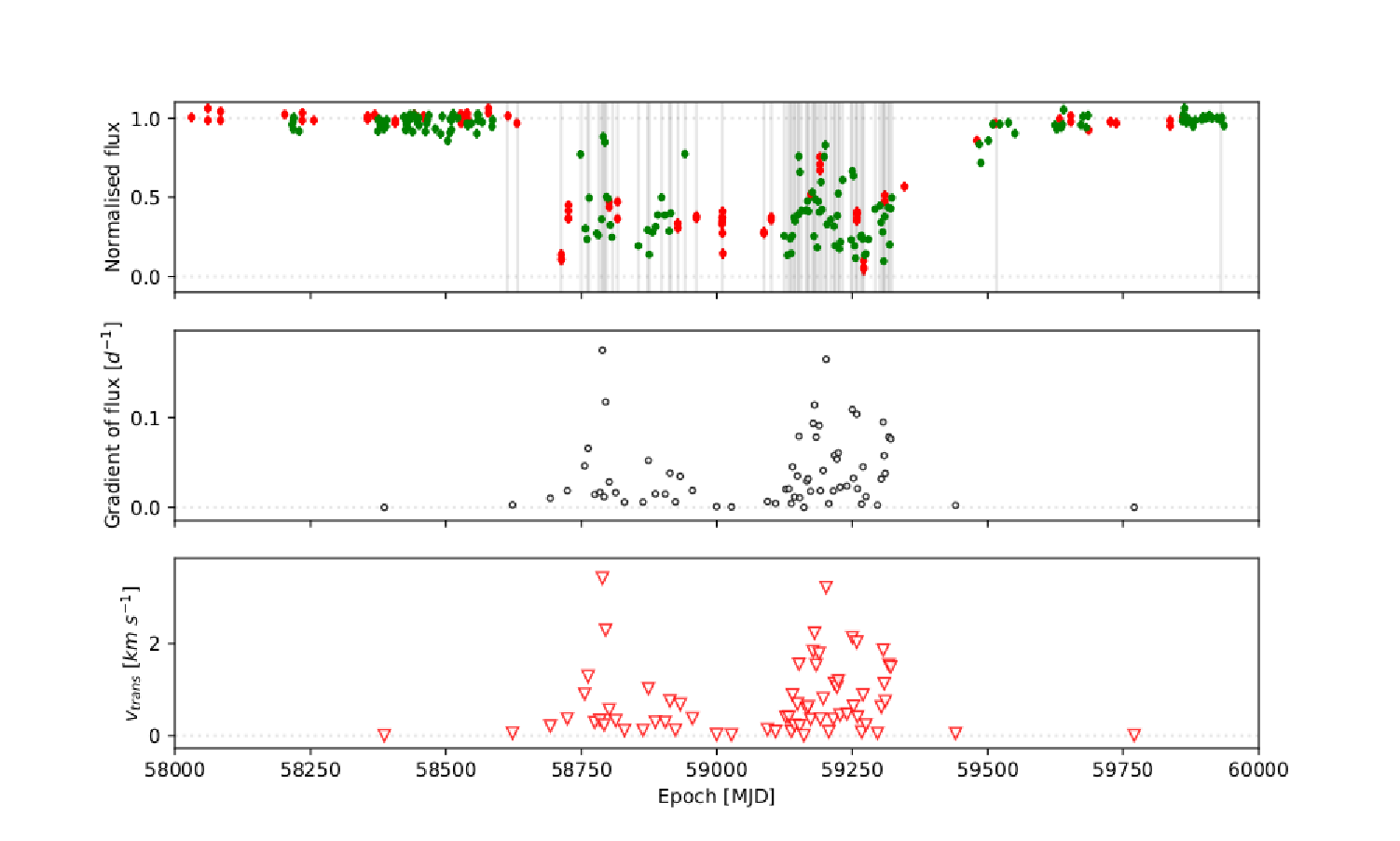}
		\caption{Analysis of the light curve gradients as a velocity of the obscuring object with the exoring analysis procedure. The analysis indicates a lower limit to the
                 transverse velocity of 2 km s$^{-1}$. This slow transverse velocity indicates a large orbital radius and period for the obscuring objects,
                 of order 30 au.
                 The upper panel shows the normalized flux light curve of the star.\\
                 The gray vertical lines mark the boundaries of sections where the light curve is linearly increasing or decreasing.\\
                 The middle panel shows the measured light curve gradient as a function of time for the eclipse.\\
                 The lower panel shows the derived transverse velocity assuming the given radius of the star quoted in the text.\\
                 Error bars are shown at $1\sigma$ confidence.}
	\end{center}
\end{figure*}


We determined an estimate of the transverse velocity of the dust by analysis of the gradients in the light curve of the star.
We followed the ''exoring'' prescription in 
\citet{vanWerkhoven.2014.MNRAS.4412845.EXORINGS}, where we obtain an 
upper bound on the transverse velocity by assuming that each of the individual flux dips 
is caused by an opaque occulter much larger than the diameter of the star that it is transiting.
When the edge of the occulter is perpendicular to the direction of motion, 
this enables an upper estimate of the transverse velocity of the cloud, 
since an edge that is not perpendicular to the vector of motion will 
require a higher velocity to cover the entire disk of the star.

The method is as follows: the light curve is converted from stellar magnitudes to flux, 
and the flux light curve is divided by the ``out of eclipse'' flux level to produce a normalized flux curve 
(see upper panel of Fig.~6).
The light curve is visually inspected to identify turning points (indicated by gray vertical lines in the upper panel) 
which we associate with a new edge with different opacity beginning to transit the stellar disk.
A straight line is then fit to the photometry between these turning points, 
and a gradient with an associated measurement error is determined - this is plotted in the middle panel of Fig.~6.
Together with the diameter of the star, a lower bound on the transverse velocity of the material can be determined.
Assuming a radius of $1.7 R_{\odot}$ for the star, the transverse velocity is shown in the lower panel of Fig.~6.

The analysis showed a robust lower speed of 2 km s$^{-1}$ for the occulter.
This lower limit is partly due to the limited sampling of the light curve during the major minima,
before the dedicated high-cadence monitoring was initiated. 
Multiplying the velocity by the duration of the eclipse gives a lower estimate for the diameter of the occulter.
For an eclipse duration of 866 days and a velocity of 2 km s$^{-1}$ the estimated diameter of the eclipsing disk is 1.0 au.
The 866 days duration includes a minor gradient before the first observed deep minimum and establishes a more
symmetric light curve.
The distribution of the gradients as a function of time show approximate time symmetry, 
with a notable dip in the measured gradients at the approximate midpoint of the eclipse.
Using the analysis of light curve gradients as a function of time as described in
\citet{Kenworthy.2015.ApJ.800.126K.J1407B.rings}
this is qualitatively consistent with a disk with azimuthally symmetric substructure 
that is moderately inclined and that the projected semi-major axis of the ring system 
is close to being parallel with respect to the path of the star behind the rings.
Assuming, for an order of magnitude estimate, that Gaia21bcv has a mass of 1 $M_\odot$, an orbital velocity of
2 km s$^{-1}$ implies a radius of a circular orbit of 225 au, and an orbital period of 3375 years.
Since the orbital velocity is a lower limit, these estimates for radius and orbital period are
upper limits. 
Even with this caveat, repeat observation of a second transit are a very distant prospect.

\subsection{Is an Orbiting Disk Consistent with the Age ?}
The SED of Gaia21bcv outside of the occultation events is essentially a reddened photosphere, with only minimal indication of
a possible infrared excess in the $W3$ filter, leading to the classification of Gaia21bcv as
a Class III star whose protostellar disk has largely dispersed. 
While there is a large scatter
in the disk clearing times for individual stars, reviewed by 
\citet{Williams.2011.ARA&A.49.67.disk.dispersion}, its Class III properties suggest an
age of at least 10 Myr for Gaia21bcv.
Together with the membership in an OB association and its pre-main-sequence luminosity,
all this points to Gaia21bcv having an age in the range of 10 - 40 Myr.
It is therefore plausible that at an age when the protoplanetary disk around the
primary star Gaia21bcv has largely dissipated, a smaller (only 1 au) disk at a
substantial distance from the primary star, orbiting a secondary, less massive
component, has escaped disk clearing and may still persist. 
While it appears clear that a massive body is needed
to keep that occulting disk together, this object appears to be invisible and we cannot constrain it mass.
Anything from a low mass star, a future brown dwarf or a young giant planet would
be possible. It may be possible to observe the thermal
emission from the occulting disk to place constraints on its internal heating by the secondary object
and thereby on the luminosity of that object.

\section{Summary and Conclusions}

Gaia21bcv has undergone an episode of repetitive, deep minima in its brightness 
between 
2019 Aug. 18 (MJD 58713)
and 2021 Nov. 2 (MJD 59520),
after showing constant brightness before and after this dipping episode.
The star is a young, probably still weakly accreting pre-main-sequence star and most likely
a member of the CMa OB1/R1 association. 
The star is approximately of K4-5 spectral type with strong metal absorption lines, including
Li absorption that indicates its youth. H$\alpha$ was weakly detected in absorption and
was variable. Variable [\ion{O}{1}] emission was detected after the end of the dipping episode and extended [\ion{S}{2}] was detected, both indicating shock-excited gas.

The dipping episode can be understood in a model similar to that of $\epsilon$ Aurigae: Occultation by a
large dust cloud orbiting the primary star. The dust cloud is probably surrounding a star or planet sufficiently massive to prevents is rapid dissipation.
In contrast to $\epsilon$ Aurigae, the occultation minimum is not as stable but consists of multiple
dipping events, suggesting
a more clumpy distribution of the dusty material or a system of rings. 
We suggest that the occulting object
may be a circumstellar or circumplanetary debris disk around 
an otherwise undetected companion object to Gaia21bcv.

\begin{acknowledgments}
    
This work has made use of data from the ESA mission
{\it Gaia}
\footnote[4]{\url{https://www.cosmos.esa.int/gaia}}
and processed by the {\it Gaia}
Data Processing and Analysis Consortium (DPAC,
\footnote[5]{\url{https://www.cosmos.esa.int/web/gaia/dpac/consortium}}
and the Photometric Science Alerts Team.
\footnote[6] {\url{http://gsaweb.ast.cam.ac.uk/alerts}}
Funding for the DPAC
has been provided by national institutions, in particular the institutions
participating in the {\it Gaia} Multilateral Agreement.\\

This work makes use of observations from the 1-m telescopes of the Las Cumbres Observatory global telescope network.  
E.G. acknowledges support from NSF Astronomy \& Astrophysics Research Grant No. 2106927.
The high-resolution spectroscopy presented herein was obtained at the W. M.
Keck Observatory, which is operated as a scientific partnership
among the California Institute of Technology, the University
of California and NASA. The Observatory was made possible
by the generous financial support of the W. M. Keck Foundation.
This publication makes use of data products from the Near-Earth Object Wide-field Infrared Survey Explorer (NEOWISE), 
which is a joint project of the Jet Propulsion Laboratory/California Institute of Technology and the University of Arizona. 
NEOWISE is funded by the National Aeronautics and Space Administration.\\
The light curve is partly based on observations obtained with the 
Samuel Oschin 48-inch Telescope at the Palomar Observatory 
as part of the Zwicky Transient Facility project. 
ZTF is supported by the National Science Foundation under Grant No. AST-1440341 
and a collaboration including Caltech, IPAC, the Weizmann Institute for Science, 
the Oskar Klein Center at Stockholm University, the University of Maryland, 
the University of Washington, Deutsches Elektronen-Synchrotron and Humboldt University,  
Los Alamos National Laboratories, the TANGO Consortium of Taiwan, the University of Wisconsin at Milwaukee, 
and Lawrence Berkeley National Laboratories. Operations are conducted by COO, IPAC, and UW.
This work is based in part on
near-infrared imaging data from the WFCAM at the UKIRT observatory operated by the University of Hawaii.
This publication makes use of data products from the Two Micron All Sky Survey, which is
a joint project of the University of Massachusetts and IPAC/Caltech, funded by NASA and NSF.\\

We thank A. M. Boesgaard for helpful discussions and W. Varricatt for help with
the UKIRT observations.

\end{acknowledgments}\vspace{5mm}

\facilities{Gaia, WISE, Keck:I, UKIRT, UH:2.2m, LCO}



\end{document}